\documentstyle[12pt,prbbib,aps]{revtex}

\begin{document}
\title{
Freezing transition in three and two dimensions by the
generalized density functional theory}
\author{ E. E. Tareyeva and V. N. Ryzhov}
\address{Verecshagin
Institute for High Pressure Physics, Russian Academy of Sciences, Troitsk
142092, Moscow region, Russia
}
\maketitle

Freezing is one of the most universal phenomena in nature.
The ability to crystallize is common to quite different
systems: systems of simpliest atoms as well as of large
molecules of complicated shape, systems of neutral as well
as of charged particles, artificial colloidal suspensions of
polysterene spheres as well as ionic plasma in white dwarfs.

The freezing transition may be induced either by
cooling or by compression, and the problem of measuring
melting curves is a classic one,
having long standing history. To--day, due to the
diamond anvil technique some melting curves are known with
great accuracy and up to Megabar pressures.
Nevertheless, the detailed microscopic mechanisms even of the
classical phenomenon of 3D melting are far from being
understood.

There are new problems which gained much interest during last
decades connected with the problem of non periodic
intermediate or ground states.  A system which under normal
conditions freezes into a crystalline phase can be made to
freeze into either a quasicrystalline or glassy state by a
suitable change in the rate of cooling. Further, there are
systems in nature which do not transform directly from isotropic
liquid to crystalline solid but exhibit a number of intermediate
phases.  Among those are surely some
2D systems and liquid crystals.

It is interesting to notice that although the window glass is
widely used for a very long time, a great number of
experimental investigations on glasses has been realized just
during last decade. Two discoveries seem me the most fascinating
among them \cite{Sc}. First, the fact that the most part of
water in the Universe occurs to be in the glass state, being
condensed from the gaseous phase at very low temperatures.
Second, it is just the transition to a glassy state and the
slowing of diffusion processes that provide the possibility of
alimentary product conservation and the life continuation in the
desert.

There is no general theoretical
argument that thermodynamically stable solid states must have a periodic
density distribution (see \cite{ruel82}).
Evidence for the linear stability of aperiodic
packing to small few-particle displacements can be obtained from a
self-consistent phonon theory \cite{stwo84}.

In the density-- functional theory
(DFT) by Singh et al. \cite{sistwo85}
to the glass transition a first-order
transition at the negative pressure was obtained.
In this connection the papers by Stishov \cite{sti1,sti2} on phase
transitions in expanded matter should be mentioned.
Interesting indications can be found also in the theory of spin systems.

The study of disordered spin systems has become a rich and
productive science. It seemed some time ago that attention
completely shifted to neural networks. However, quite recently
a number of
papers has appeared concerning the possilility of obtaining a kind of
spin glass regime in initially non-random systems.
Glasses, of course, have no disorder in their Hamiltonian -- the
randomness is formed in the transition. The deterministic spin
models with self-induced quenched disorder
represent a convenient
candidate to describe real glasses.
Great expectations can be connected with the using of the methods of
spin glass theory in the theory of real glasses,
particularly if combined with DFT of freezing.

There is another new aspect of the problem in question, which is
connected with a
new concept which appears in the physics of liquid -- solid transition
just before 1980: bond orientational ordering.
If we consider crystallization as a broken symmetry transition
we can see that two distinct broken symmetries distinguish
crystalline solid from isotropic liquids: translational and
rotational ones. These two symmetries are not independent,
because rotating one patch of perfect crystal relative to
another clearly disrupts not only orientational correlations,
but translational correlations as well. A relative translation of the two
patches, on the other hand, decorrelates translational order, but leaves
orientational correlations intact. In such a way one can obtain
an exotic fluid or glass state of
matter with extended correlations in the orientations of
locally-defined crystallographic axes, but with short-range translational
order. This subtle type of order is called bond-orientational order (BOO)
(a "bond" joining near-neighbor atoms)
Anisotropic
fluids of this kind are a part of recent theories of 2D melting
\cite{strandburg88} proposed by Halperin and Nelson
\cite{halpnel79} and Young \cite{young} and based on the ideas of
Kosterlitz and Thouless \cite{kosthoul73}.  This KTHNY theory
predicts that the transition may be fundamentally different from
that observed in ordinary three-dimensional systems. It was
found that the transition between two-dimensional solid and
isotropic liquid can occur via two continuous transitions
corresponding, respectively, to dissociation of dislocations and
disclinations. The intermediate phase with BOO was called the
hexatic phase. The properties of the hexatic phase are similar
to those of a nematic liquid crystal, but with a cluster of
spherical atoms instead of extended molecules.
This theory has strong support from experiments with electrons on
helium, computer simulations of 2D electron systems,
experiments with polystyrene spheres and others.
However, a conventional
first-order transition between the two-dimensional solid and
isotropic liquid is also a possibility.

There are some indications that BOO can exist in three-dimensional fluids
and glasses, too (Steinhardt, Nelson, Toner, Hess, Mitus
and Patashinskii, Ryzhov). It should be emphasized, however,
that there is a great difference between 2D and 3D cases. In two
dimensions the symmetry of elementary clusters is
crystallographic so that they can simply order in a crystal.
In three dimensions the energetically preferable symmetry
of clusters for standard potentials
is usually icosahedral \cite{Frank}, incompatible with 3D
crystal. In the frame of the fivefold symmetry one
can pack only 96 (not infinite number) particles. This means
that the cooling of three-dimensional analog of hexatic phase
can not give the icosahedrons packing into regular
three--dimensional crystal:
 the clusters must be destroyed
 before the crystallization takes place.
 It is easier to obtain such a
 crystal from usual (not supercooled) liquids, where there are
 crystallographic -- hexagonal and cubic -- clusters. As to
 supercooled anisotropic liquid with icosahedral clusters --
 it possibly freezes into a glass phase.
 We shall discuss the
 problem of BOO later in more details.

 Now it is the time to make a remark about 2D crystals. Strictly
speaking there is no 2D crystals in the thermodynamic limit:
they are destroyed by fluctuations. In 1935 Peierls
~\cite{Pa35} has shown that if the temperature is nonzero
then the longwavelength phonons destroyed long range order
in 2D crystals: the meansquare deviations of atoms
from their equilibrium positions increase logarithmically
with the size of system and the Bragg diffraction peacks become
smoothed. Later such absence of two--dimensional long--range
order was proved by Mermin ~\cite{merm68} with the use of well
known Bogoliubov inequalities ($1/q^2$ theorem) for
correlators \cite{bogol1}.  Now it is clear that one can
discuss the quasilongrange translational order in 2D systems.
The appearence of such an order means that the longrange
correlations change the character of decay from exponential to
algebraic.

In this report we shall discuss some theoretical aspects of the
phenomenon of freezing, mainly the density--functional theory of
the freezing of 3D and 2D systems. The fascinating problem of
glass formation remains beyond the scope of this report,
although there are some indications on the applicability of
DFT in this case, too. This report has no intention of
completeness. The choosen list of results and references is
influenced by the authors own experience in DFT of freezing and
in two--dimensional melting as well as by fruitful discussions
with high-pressure physisists.

A great deal of information on the phenomenon of freezing
can be found in reviews (see, for example, the reviews Ref.
\cite{sti74,sti88,sing91} and Les Houches sessions
\cite{leho88,leho89} of 1988 and 1989  for 3D systems and
\cite{stran88,stran92} for the 2D case ).

I shall begin with some very simple facts which are not widely
known because the crystallization problems are usually avoided in
courses and in standard manuals and textbooks.

The crystallization transition is strongly first-order in three-dimensions
and is marked by large discontinuities in entropy, density and order
parameters.  The correlation functions remain
short-ranged near the transition. In spite of this the transition is known
to have a kind of quasiuniversal behaviour. One can mention the following
"universalities".
\begin{enumerate}
\item The Lindemann criterion states that  the crystal melts
when the ratio of the
root-mean-squared thermal vibration amplitude to the nearest-neighbour
spacing attains a certain critical value ($\sim 0.1$).
The Lindemann criterion has been found to be reasonably correct for a
number of simple systems. From computer simulations and indirectly, from
experiments, the ratio is indeed seen to be about 0.09 for close packed
fcc structures and about 0.12 for open bcc structures.
This is rather close to the actual
melting point and gives a quasi-universal criterion of melting.

\item A comparable one-phase criterion, known as the Hansen-Verlet criterion
\cite{have69}, exists on the fluid side of the coexistence
curve. During a course of extensive computer simulations of
simple fluids, Hansen and Verlet found that the amplitude of the
main peak of the structure factor $S(k_m)$ is approximately
constant along the crystallization line.  There is direct
experimental evidence that simple fluids as different as sodium
and argon, in addition to model systems such as a hard-spheres
fluid, the LJ fluid, one--component plasma (OCP) etc., all
freeze when $S(k_m) \simeq 2.9 \pm 0.1$.

\item There is a kind of universality in the form of melting curves.
Except some exotic curves with maxima
in p-T coordinates they are monotonic and have no
critical end points. There are three variants of
tricritical points : Liquid-Gas-Solid; LiduidI-LiquidII-Solid;
Liquid-SolidI-SolidII. The examples of calculations of such
curves with tricritical points
can be found in the recent papers by Baus and coworkers
\cite{baus1,baus2,baus3}
where the role of repulsive and attractive parts
of pair potential in crystallization is analyzed in detail.

\item There is a kind of universality in the behaviour of thermodynamic
functions along melting curves. It seems that the entropy change on
melting is constant for a given material and that there is the following
trend \cite{sti74,sti88}:
$$
\frac{\Delta S}{R} \to \ln 2 ~ \mbox{when} ~
 \frac {\Delta V} {V_s} \to 0.$$

\item Closely connected with the previous "universality" is a kind of
well known "scaling" observed in the course of detailed investigation
of the freezing of soft spheres systems with inverse power
potential:  $$u(r)=\varepsilon(\sigma/r)^n,$$ where
$\varepsilon$ and $\sigma$ measure the strength and the
characteristic length of the interaction.  The reduced excess
thermodynamic properties of the soft spheres depend on a single variable
which is defined as $$\gamma =(\varrho
\sigma^3)(\varepsilon/{k_{B}T})^{3/n} = \varrho^* T^{* -3/n}.$$
Freezing of soft-spheres fluids have been
extensively studied by MC simulation for several values of the exponent
$n$, e.g., $n=12$ \cite{hans70,horojo70}, 9, 6 and 4
\cite{hogrjo71,hasc73}. The two extreme cases are:
the HS system ($n=\infty$), and one component plasma (OCP)
($n=1$).
The computer
simulations have revealed some symmetric trends in the melting
properties of soft spheres when $n$ decreases from $\infty$ to OCP
\cite{horo71,hoyogr72}.
The relative volume change on melting decreases
rapidly with $n$, while the entropy change per particle on
melting is relatively insensitive to $n$.  Moreover, the more
repulsive systems ($n\geq 7$) freeze into a close-packed fcc
structure, while the soft repulsions ($n\leq 7$) lead to
crystallization into a bcc phase. The Lindemann and the
Hansen-Verlet criterion are exactly obeyed for any $n$.  There
is also rather general result by Weeks \cite{week81} who has
shown for the case of arbitrary dimensions that for systems
interacting by means of purely repulsive power--law potentials,
$\sim r^{-m}$, the change in specific volume $\Delta v$ on
melting approaches zero as $m \to d$, where $d$ is the
dimensionality of the system.

\end{enumerate}

All these facts are pointing towards the possibility of existing an
underlying theory. But, in fact, there is no such theory so far.
It should be emphasized that although the
freezing is an example of phase transition but different theoretical
methods employed for the study of other types of phase transitions cannot
be employed in this case:
1) Many phase transition theories are based on lattice models: many
magnetic transitions, gas-liquid transition. For the
liquid-crystal transition lattice models have not proved very
successful because the periodicity imposed by the lattice cannot
be separated from the spontaneous order that ought to arise upon
crystallization.  2) The renormalization group (RG) approach has
not found much use for studying first-order transitions.  3)
Computer simulations have given a great deal of information, but
again have limitations. The system sizes that can be simulated
are small, and the time scales are short. The question always
remains whether the time scale
of the simulation is sufficient to reach a true equilibrium between two
phases.
That is why density functional theories (DFT) of freezing occur
to be rather useful although they have their own limitations,
too. DFT is "good" particularly for hard core potentials and for high
densities because it is a geometrical theory based on packing
picture entering the theory through direct correlation function (DCF).
However the conventional DFT occurs to fail in the case of long range
potentials and can not describe the
melting of some two--dimensional systems.

Now we shall briefly recall the main points of the conventional DFT
of freezing, list some of the most interesting recent results in DFT and
discuss our results on the amelioration of this theory. Particularly,
we shall reformulate the basic equations of the DFT
as to describe 2D melting in terms of
distribution and correlation functions.
The resulting integral equations theory can be considered as a variant
of DFT theory appropriate to 2D melting and, possibly, glass transition.

The physical idea behind the DFT is the fact that at the freezing
transition the correlation length is only a few atomic spacings. All
phenomena at distances greater than the correlation length can be treated
in a mean-field approximation.
Implicit in this approach is an
assumption according to which a system is either entirely in the liquid
or entirely in the ordered phase, where no phase coexistence is
permitted. Fluctuations are thought to be not of great importance in
a completely entropy-driven first-order transition.

The DFT approach to crystallization is based on the theorem \cite{evan79}
that the Helmholtz free energy $F[\varrho({\bf r})]$ of an inhomogeneous
system is a {\it unique functional} of the one--particle density
$\varrho ({\bf r})$, which in a crystalline solid is extremely
inhomogeneous.
The mathematics of DFT is the bifurcation theory for the
solutions of nonlinear integral equations for one--particle
distribution function. This idea first appeared in the
papers by Kirkwood and Monroe \cite{kimo}, Tyablikov \cite{tyab}
and Vlasov \cite{vlas}.

The first papers on the DFT approach to crystallization problem
appeared in 1979--1981 \cite{rayu79,ryta79,ryta81,haymet81}.

To describe freezing in three (and two) dimensions we shall
follow our papers \cite{ryta79,ryta81,ryz83,rytaph} and use
a new formalism -- that of classical
many--particle conditional distribution functions.
These functions $F({\bf r}_1|\Psi)$ give the
probability of finding a particle at ${\bf r}_1$
in the external field $\Psi({\bf r}_1)$
The equations for these functions
can be obtained from
the non-linear integral equation for the singlet distribution function
$F({\bf r}_1|\Psi)$ in an external field $ \Psi({\bf r}_1)$:
\begin{eqnarray}
 \rho \frac{F({\bf r}_1|\Psi)}{z}&  = & \exp
\left\{ -\beta \Psi({\bf r}_1)+ +\sum_{k \geq 1} \frac{\rho^k}{k!} \int \,
S_{k+1}({\bf r}_1,...,{\bf r}_{k+1}) \right.  \nonumber \\ & &  \times
\left. F({\bf r}_2 | \Psi) \cdots F({\bf r}_{k+1}|\Psi)\, d{\bf r}_2... d{\bf
r}_{k+1} \right\}
\label{rho},
\end{eqnarray}
Here
$z $ is the activity, $\rho$ is the mean number density, $S_{k+1}({\bf
r}_1,...,{\bf r}_{k+1})$ is the irreducible cluster sum of Mayer functions
connecting (at least doubly) $k+1$ particles,
$\beta=1/k_B T$ and $T$ is the temperature.

     This equation was derived for canonical ensemble by
Arinstein \cite{arin} on the base of Bogoliubov functional
method \cite{NNB2} and then rederived (for the case of small
$\rho$) by Stillinger and Buff \cite{stilbuf} who have used the
diagram technique. The simpliest way to derive it in the case
of grand canonical ensemble was given by Ryzhov
\cite{thes}.

If the external field has the form
$$ \Psi({\bf r}_1)= \sum_{k+1}^s \Phi({\bf r}_1-{\bf r}_k^0) $$
where $\Phi(r)$ is the interparticle potential, then the function
$F({\bf r}_1|\Psi)$  is the probability of finding a particle at
${\bf r}_1$, if $s$ particles are at the points
${\bf r}_1^0,...,{\bf r}_s^0$
\[ F({\bf r}_1|\Psi)=F_{s+1}({\bf r}_1|{\bf r}_1^0 ... {\bf r}_s^0)=
\frac{F_{s+1}({\bf r}_1, {\bf r}_1^0,...,{\bf r}_s^0)}
{F_s({\bf r}_1^0,...,{\bf r}_s^0)}. \]
Here $F_s({\bf r}_1,...,{\bf r}_s)$ is the $s$--particles distribution
function.

In this case the equation (\ref{rho}) takes the form
\begin{eqnarray}
\frac{\rho F_{s+1}({\bf r}_1|{\bf r}_1^0 ... {\bf r}_s^0)}{z}& = &\exp \left\{
-\beta \sum_{k+1}^s \Phi({\bf r}_1-{\bf r}_k^0)
+\sum_{k \geq 1} \frac{\rho^k}{k!} \int \,
S_{k+1}({\bf r}_2,...,{\bf r}_{k+1}) \right. \nonumber \\
& &\left.\times F_{s+1}({\bf r}_1|{\bf r}_1^0 ... {\bf r}_s^0)...
F_{s+1}({\bf r}_{k+1}|{\bf r}_1^0 ... {\bf r}_s^0)
d{\bf r}_2... d{\bf r}_{k+1} \right\} \label{main}.
\end{eqnarray}

The value of $z$ in general case can be obtained from the normalization
condition
\begin{equation}
 \frac{1}{V} \int \,F_{s+1}({\bf r}_1|{\bf r}_1^0 ... {\bf r}_s^0)\,
d{\bf r}_1 =1.
\label{nor}
\end{equation}

If one takes the derivative of
(\ref{main}) relative to ${\bf r}_1$, one obtains the equilibrium BBGKY
hierarchy
\cite{NNB2}
\begin{eqnarray}
 k_B T \nabla_1 F_{s+1}({\bf r}_1, {\bf r}_1^0,...,{\bf r}_s^0) & +
& F_{s+1}({\bf r}_1, {\bf r}_1^0,...,{\bf r}_s^0) \nabla_1
\sum_{k+1}^s \Phi({\bf r}_1-{\bf r}_k^0) + \nonumber\\
 & + & \rho \int \nabla_1 \Phi({\bf r}_1-{\bf r}_2)
F_{s+2}({\bf r}_1, {\bf r}_2, {\bf r}_1^0,...,{\bf r}_s^0)\,
d{\bf r}_2 = 0,\nonumber
\end{eqnarray}
along with the explicit expression for
$F_{s+2}$ as the functional on $F_{s+1}$ :
\begin{eqnarray}
F_{s+2}({\bf r}_1, {\bf r}_2, {\bf r}_1^0,...,{\bf r}_s^0) & = &
F_{s}({\bf r}_1^0,...,{\bf r}_s^0) e^{-\beta \Phi({\bf r}_1-{\bf r}_2)}
F_{s+1}({\bf r}_1|{\bf r}_1^0 ... {\bf r}_s^0)
F_{s+1}({\bf r}_2|{\bf r}_1^0 ... {\bf r}_s^0) \nonumber \\
& \times & \sum_{k \geq 1} \frac{\rho^{k-1}}{(k-1)!}\, \int\,
\frac{\partial S_{k+1}({\bf r}_1,...,{\bf r}_{k+1})}{\partial f(r_{12})}\,
F_{s+1}({\bf r}_3|{\bf r}_1^0 ... {\bf r}_s^0) ... \nonumber \\
&...& F_{s+1}({\bf r}_{k+1}|{\bf r}_1^0 ... {\bf r}_s^0) d{\bf r}_3... d{\bf r}_{k+1},
\label{add}
\end{eqnarray}
with
\[ f(r_{12})=e^{-\beta \Phi(r_{12})}-1 \]

This equation gives the exact closure. However it contains
infinite series and integrals and one has to use some
approximations to exploit it. The same can be said about the
Eq.(\ref{main}) itself. It is formally closed although the gain
is not obvious: the price is the infiniteness of series.

The free energy functional (FEF) of such inhomogeneous system
with the density $\rho({\bf r}) = \rho F_1(\bf r)$ has the form:
\begin{equation} \begin{array}{lcl}
{\cal F}/k_BT = \int\,d{\bf r}_1\,
\rho({\bf r}_1)[\ln(\lambda^d\rho({\bf r}_1)-1]-\\
 - \sum_{k \geq 1}
\frac{1}{(k+1)!} \int \cdots
\int\, S_{k+1}({\bf r}_1...{\bf r}_{k+1})\rho({\bf r}_1) \cdots
\rho({\bf r}_{k+1})\,d{\bf r}_1 \cdots d{\bf r}_{k+1}.\end{array}
\label{free}
\end{equation}
or
\begin{equation}
{\cal F}/k_BT = \int\,d{\bf r}_1\,
\rho({\bf r}_1)[\ln(\lambda^d\rho({\bf r}_1)-1]-{\cal
F}_{ex}[\rho({\bf r})]/ k_BT.
\label{frex}
\end{equation}
The excess
free energy ${\cal F}_{ex}[\rho({\bf r})]/k_BT$ is just the
generating functional for direct correlation functions
\begin{equation}
c_n({\bf r}_1...{\bf r}_n)=\frac{\delta^n {\cal
F}_{ex}[\rho({\bf r})]/k_BT} {\delta \rho({\bf r}_1) \cdots
\rho({\bf r}_n)}.
\end{equation}

If the external field is the potential of atom held fixed at the origin
then the functional to be extremized depends only
on the one particle distribution function $\varrho ({\bf r})$
and we can write the Taylor expansion for the excess free
energy around the liquid in the following form:
\begin{equation}
\beta \Delta F = \int d{\bf r} \varrho ({\bf r}) \ln \frac {\varrho ({\bf
r})} {\varrho _0} - \sum_{k \geq 2} {1 \over k!} \int
c^{(n)} ({\bf r}_1,...,{\bf r}_k)
\Delta \varrho ({\bf r}_1)...\Delta\varrho ({\bf r}_k)
d{\bf r}_1 ... d{\bf r}_k ,
\label{exfree}
\end{equation}
where
$$ \Delta \varrho ({\bf r}) = \varrho ({\bf r}) - \varrho_l $$
is the local density difference between solid and liquid phase.

Integral equation for $\Delta \varrho ({\bf r})$ which
extremizes $\Delta F$ is formally closed and nonlinear. The
bifurcation point of the trivial solution of this equation
determines temperature and pressure when the nontrivial solution
appears. This is the point of absolute instability of liquid
phase against nonconstant density state formation. It can be
obtained through exact linearisation of the Eq.(\ref{exfree}).
We can obtain
\begin{equation}
\nabla_1
\ln \varrho({\bf r}_1)=\int\,d{\bf r}_2 c_2({\bf r}_1,{\bf r}_2)
\nabla_2 \varrho({\bf r}_2).
\label{lov}
\end{equation}
or, for the Fourier--transforms:
\begin{equation}
\delta \varrho(k)=-\beta\rho S(k)\tilde{U}(k),
\end{equation}
где $S(k)=1+\rho\int\,d{\bf r}e^{i{\bf kr}}[g(r)-1]$.
$S(k)$ has the following form in terms of DCF:
\begin{equation}
S(k)=\frac{1}{1-\rho \tilde{c}_2(k)}
\label{strf}
\end{equation}
Now one can see that the poles of the structure factor
$S(k)$ (the instability points) given by the Eq.
(\ref{strf}) are the points of mechanical instability of the
system.

In order to obtain the actual point
of the phase transition one must use the conditions of equality
of chemical potentials and pressures for liquid and solid phases
and to compare the free energy values.

The full system of equations to be solved in the DFT contains
the nonlinear integral equation for the function $\rho ({\bf
r})$, obtained as the extremum condition for the excess free
energy and the equilibrium conditions
\begin{equation}
(T_1=T_2);\,\,\,\,P_1=P_2;\,\,\,\,\mu_1=\mu_2; 
\end{equation}
 with $\mu_i$ - the chemical potential and $P_i$ - the pressure
of the $i$-th phase, written in terms of the same functions as
in (\ref{exfree}). Namely the chemical potential of the
inhomogeneous phase is
\begin{equation}
\mu=\frac{1}{V}\int\,d{\bf r}\frac{\delta{\cal F}(\rho({\bf r}))}{\delta
\rho({\bf r})} 
\end{equation}

 The pressure in the solid phase is
\begin{equation}
P_s=P_l+\sum_{n=1}\frac{1}{n!}\int\,
\left. \frac{\delta^n P}{\delta\rho({\bf r}_1) \cdots \delta\rho({\bf r}_n)}
\right|_{\rho({\bf r})=\rho_l}
\Delta\rho({\bf r}_1) \cdots
\Delta\rho({\bf r}_n)\, d{\bf r}_1 \cdots d{\bf r}_n. 
\end{equation}
 where $P$ is the functional of
 $\rho({\bf r})$ :
\begin{equation}
P=\frac{k_B T}{V}\left\{\int\,d{\bf r}\rho({\bf r})(1-
c_1({\bf r};\{\rho({\bf r})\})+
\beta{\cal F}_{ex}(\rho({\bf r}))\right\}. 
\end{equation}

 To proceed constructively in the frame of the DFT we must
choose a concrete form of FEF -- a kind of closure or
truncating -- and we must make an ansatz for the average
density of the crystal. The importance of such ansatz follows
from the fact that we are dealing with a theory
which is equivalent to Gibbs distribution and one has to
break symmetry following the Bogoliubov concept of
quasiaverages \cite{bogol1}. Now it is necessary to specify
the crystal symmetry (e.g.lattice type) and to locate
the freezing transition for that particular lattice type. Other lattice
types can then be studied as well. There is always the possibility,
though, that a more complex lattice that has not been examined will turn
out to be more stable.

Let us demonstrate now how the DFT works in the simpliest
case of hard sphere system truncating the free energy
functional in the
hypernetted chain (HNC) approximation.
The one particle distribution function
$\rho({\bf r})$ in the crystalline phase has the form
\begin{equation}
\rho({\bf r})=\rho_l+\Delta\rho({\bf r}),\,\,\,\,\frac{1}{V}\int\,
\rho({\bf r})\,d{\bf r}=\rho_s.     
\end{equation}
Here $\rho_l$ and $\rho_s$ - are the averaged density of liquid
 and solid phase, respectively, and the function
 $\Delta\rho({\bf r})$  contains the term having the crystal
 symmetry.
Let us use the following actual form for the
density change
  $\Delta\rho({\bf r})$ :
\begin{equation} \begin{array}{rcl}
\Delta\rho({\bf r})&=&\rho_l\sum_{{\bf k}}\varphi_{{\bf k}}e^{i{\bf kr}}=
\rho_l\varphi_0+\rho_l\varphi({\bf r}),\\
\varphi_{{\bf k}}&=&\frac{1}{\Delta}\int_{\Delta}\,\frac{\Delta\rho({\bf r})}
{\rho_l}e^{-i{\bf kr}}d{\bf r}.\end{array}     \label{233}
\end{equation}
The sum is over reciprocal lattice vectors and the integral is
taken over the elementary lattice cell
$\Delta$ .

If $J$ labels the sets of the points in reciprocal lattice
with equal coefficients $\varphi_{{\bf k}}$ (from symmetry
condition), then
\begin{equation} \Delta\rho({\bf r})=
\rho_l\varphi_0+\rho_l\sum_J\varphi_J\xi_J({\bf r}),
\,\,\,\,\xi_J({\bf r})=\sum_{{\bf k}\in J}e^{i{\bf kr}}. 
\end{equation}
and $\varphi_J({\bf r})$ are independent order parameters.

In the HNC approximation
$c_n({\bf r}_1 \cdots {\bf r}_n)=0, \,\,n\geq 3$,
and we obtain from $P_l = P_s$ the following equation for
the density change
$\Delta \rho({\bf r})$:
\begin{equation}
(1-\rho_l\tilde{c}_2(0))\,\int\,\Delta
\rho({\bf r}) d{\bf r}-
\frac{1}{2}\,\int\,c_2(|{\bf r}_1-{\bf r}_2|)\Delta
\rho({\bf r}_1)
\Delta \rho({\bf r}_2) d{\bf r}_1 d{\bf r}_2 = 0,
\label{derho}
\end{equation}
or
\begin{equation}
(1-\rho_l\tilde{c}_2(0))\varphi_0
\rho_l-\frac{1}{2}\varphi_0^2\rho_l^2
\tilde{c}_2(0)-\frac{1}{2}\rho_l^2\sum_J m_J \tilde{c}_2(k_J)
\varphi_J^2=0,
\label{derhoj}
\end{equation}
where
\begin{eqnarray}
m_J&=&\frac{1}{V}\int\,\xi_J^2({\bf r}) d{\bf r},\\
\tilde{c}_2(k)&=&\int\,c_2(r) e^{-i{\bf kr}} d{\bf r}.
\label{derhj1}
\end{eqnarray}

The equality of chemical potentials gives the equation
\begin{equation}
(1+\varphi_0)\exp(-\rho_l\varphi_0\tilde{c}_2(0))=
\frac{1}{V}\int\,
d{\bf r}\,\exp\left\{\rho_l\sum_J\tilde{c}_2(k_J)
\varphi_J\xi_J({\bf r})
\right\}.
\label{244}
\end{equation}
Finally, using the equation
\[\frac{1}{V}\int\,\xi_J({\bf r})\xi_{J'}({\bf r}) d{\bf r}=m_J\delta_{JJ'},\]
it is easy to obtain
\begin{equation}
m_J\frac{\varphi_J}{1+\varphi_0}=\frac{
1/V\int\,d{\bf r}\xi_J({\bf r})\exp\left\{
\rho_l\sum_J \tilde{c}_2(k_J)
\varphi_J\xi_J({\bf r})\right\}
}{
1/V\int\,d{\bf r}\exp\left\{\rho_l\sum_J \tilde{c}_2(k_J)
\varphi_J\xi_J({\bf r})\right\}}.
\label{245}
\end{equation}

The equations (\ref{derhoj}), (\ref{244})  и (\ref{245})
present the closed system of equations for the quantities
$\rho_l, \varphi_0, \varphi_J$ -- the liquid density at the
transition, the density change at the crystallization and the
components of the one--particle distribution function,
respectively.

One can also obtain the entropy change in the form
\begin{equation}
\frac{\Delta
s}{k_B}=-\left((\varphi_0-\varphi_0^2)\frac{P_l}{\rho_l k_B
T} +\varphi_0^2(1-\rho_l \tilde{c}_2(0))\right).
\label{249}
\end{equation}
Even this simple approach gives rather good results. Here in
the table we list the result of the calculation
for HS system in the case of
FCC lattice with two reciprocal lattice vector values:
$\varphi_1$ and $\varphi_2$.
\bigskip

\begin{center}
{\large Таблица 3}\\
\bigskip

\begin{tabular}{ccccc}
&$\hspace{1cm}\eta_l\hspace{1cm}$&$\hspace{1cm}\Delta\rho/\rho_l\hspace{1cm}$
&$\hspace{1cm}P^*\hspace{1cm}$&$\hspace{1cm}\Delta s/k_B\hspace{1cm}$\\
Simulations (MC)& 0.494 & 0.103 & 11.7 & 1.16 \\
This approach& 0.494 & 0.074 & 12.6 & 1.29 \\
GELA & 0.495 & 0.101 & 11.9 & 1.15\\
SCELA & 0.508 & 0.105 & 13.3 & 1.27\\
WDA & 0.480 & 0.141 & 10.4 & 1.41\\
MWDA & 0.476 & 0.139 & 10.1 & 1.35\\
ELA & 0.520 & 0.090 & 16.1 & 1.36\\
\tableline\tableline
\end{tabular}
\end{center}
\bigskip

It is worth to notice that the interparticle potential
(as well as the temperature) does not enter the equations.
This means, in particular, that they are
valid for many--body forces, too. The liquid--phase
properties enter the theory through the Fourier--transform of
the DCF, connected with the liquid structure factor $S(k)$
($S(k)=1+\rho\int\,d{\bf r}e^{i{\bf kr}}[g(r)-1]$) by
\begin{equation}
S(k)=\frac{1}{1-\rho \tilde{c}_2(k)}
\end{equation}
The structure factor characterizes the relative position
of particles in the liquid and so we can speak about the DFT
approach as a geometrical one and of packing character.

Let us emphasize that we have
truncated the Taylor expansion at second order. Although
this is a rather crude approximation it
is reasonably successful in studying  the freezing of a number
of systems.

In work done so far using this perturbation theory  either
$c^{(3)}({\bf r}_1,{\bf r}_2,{\bf r}_3)$ has been ignored
completely, or only its long-wavelength behaviour
has been included, to account for the
density dependence of the compressibility.

Although the truncated perturbation approach has significant limitations
it has one advantage over all
approaches proposed to date. This is the fact that it depends only on the
structure factor of the liquid at a fixed density in the liquid phase
region. This is experimentally measurable quantity, and this
opens up the possibility of studying the freezing of rather
complex liquids, for which the interaction potentials are not
very well known, provided that the experimental data on the
structure factor exists.

If the potential is well known for the given real system or if we are
dealing with a system subject to computer simulations then the direct
correlation function to input is usually taken from
hypernetted chain (HNC) or Percus-Yevic (PY)
approximation
 or from the interpolation closure relation
of Rogers
and Young \cite{royo84},
\begin{equation}
g(r)=\exp[-\beta u(r)](1+\{\exp[p(r)f(r)]-1\}/f(r)),
\label{rogy}
\end{equation}
where $p(r)=h(r)-c(r)$, $h(r)=g(r)-1$
and $f(r)$ is a "switching function" chosen to be
of the form
\begin{equation}
f(r)=1-\exp(- \kappa r)
\label{rogy1}
\end{equation}
The parameter $\kappa$ in $f(r)$ is varied until consistency is achieved
between the equations of state derived from virial and
compressibility routes. Another scheme is the modified HNC
(MHNC) scheme of Rosenfeld and Ashcroft \cite{roas79} based on
the closure \begin{equation} g(r)=\exp[- \beta u(r) + p(r) +
B(r)], \label{brid} \end{equation} where the bridge
function B(r) is assumed to be a universal function equal to
its hard sphere (HS) form $B(\eta,r)$ calculated for some
effective $\eta$; $\eta$ is adjusted to yield thermodynamic consistency.

 The major limitation of perturbation approaches is that
including even third-order term is very difficult. The natural
question is whether there exists a non-perturbative approach,
which might include important contributions from the other
terms.

The DFT views the emerging ordered phase as a grossly inhomogeneous liquid
with a rapidly varying one-particle density $\rho ({\bf r})$,
reflecting the lower symmetry of the emerging phase. Given such a
scenario, it seems unlikely in spite of its success, that the structure
of the ordered phase can be approximated by the low-order perturbation
expansion with a uniform system taken as the zeroth or unperturbed
system. Motivated by this fact a number of workers have attempted
to develop approximate but nonperturbative free energy
functional (FEF).

This is the idea behind the set of approaches which use an
imaginated, unreal liquid at an auxiliary density with some
kind of variational properties.

A very popular
variant of the nonperturbation approach
uses a uniform liquid which has
{\it different} density than $\rho_0$, so that chemical
potential of the solid is no longer equal to that of the liquid.
The simpliest version of such an
"effective--liquid approximation"(ELA) was proposed by Baus and
coworkers in \cite{baco85} and
chooses the reference liquid density in such
a way that the first reciprocal lattice vector $\vec k_1$
of solids matches the
first peak in the structure factor of the liquid. Such a reference liquid
is more dense than the liquid in coexistence with crystal at
the same temperature, and an extrapolation into the metastable
liquid phase is required. This is done using HS perturbation
theory to calculate the DCF and the structure factor over a
large range of densities.

A second set of approaches to constructing an approximate FEF
are referred to
as "weighted density approximations" (WDA). This approach was
due to Tarazona \cite{tara8485} and
Ashcroft and coworkers \cite{cuas85,deas89b}.

The authors have
constructed a FEF in such a way that the free-energy
density of an inhomogeneous system at a given point is interpreted as that
of a homogeneous system, but taken at an auxiliary density which depends
parametrically on the chosen point. The effective density is
obtained approximately by weighting the physical density over a given
point. The resultant weighted density approximation
(WDA) thus accounts by construction for
the short-ranged nonlocal effects present in a real, interacting
inhomogeneous liquid at the given point.

A simple version of the WDA [known as modified WDA (MWDA)] emerges
\cite{deas89a}
if one considers global free energy per particle instead of the
local excess free energy per particle. In this case the effective
density must be position independent.

Once a FEF has been specified and an ansatz has been made for the
crystal density (usually it is the form (\ref{233}) or a sum
over the direct lattice
\begin{equation}
\varrho({\bf r}) = \sum_{i=1}^{N} \varphi ({\bf r} - {\bf R_i})
\label{gaus}
\end{equation}
and assuming  a spherically symmetric Gaussian form for $\varphi$:
\begin{equation}
\varphi ({\bf r} - {\bf R_i}) =(\alpha/\pi)^{3/2} exp [-\alpha
({\bf r} - {\bf R_i})^2],)
\label{gau1}
\end{equation}
one has to calculate as in the simple example considered.

After 1979 hundreds of calculations were performed of the
melting curves of different systems.
There is a large number of results
obtained by the use of DFT which are in excellent agreement with
real or computer experiments.
It should be noticed that according to geometrical nature of the theory
higher is the pressure, better is the agreement.
The progress in the DFT
as describing the 3D classical melting is connected mainly
with the papers by Baus, Hansen, Ashcroft and their coworkers.
The details of slightly  different approaches and the lists of
results can be found in the reviews
\cite{rev1,rev2,rev3,rev4,rev5}.

Technically, the calculations follow the same scheme, mentioned
above.
The following generalizations should be mentioned.

1. In the case of charged mixture DFT is formulated in terms of number
density
$$\rho^{N}(r)=\rho^{+}(r) + \rho^{-}(r),$$
charge density
$$\rho^{Z}(r)=\rho^{+}(r) - \rho^{-}(r),$$
number-number DCF
$$c^{(2)}_{NN} = \frac{1}{2}[c^{(2)}_{++}(r) + c^{(2)}_{+-}(r)]$$
and charge-charge DCF
$$c^{(2)}_{ZZ} = \frac{1}{2}[c^{(2)}_{++}(r) - c^{(2)}_{+-}(r)].$$
Barrat \cite{bara87} has obtained the phase
diagram (using the second-order DFT and the Gaussian
parametrization for the solid density) for
a mixture of charged HS of the same diameter $\sigma$ and opposite charges
known as the restricted primitive model (RPM) for ionic liquids.

2. Freezing of molecular fluids is described in DFT modified to include
the angular dependence of potentials and DCF of the molecular liquid
and to look for angle depending solutions for the density. The
number of order parameters increases but using of symmetry
conditions reduces this number.
The number of order parameters is
determined indirectly by the number of harmonics needed to get
proper convergence, and this number increases with the
anisotropy in the shape of molecules.  The new phase transition
is the orientational phase transition. If the anisotropic part
of the potential is large enough and the shape of molecules
differs strongly from the spherical, one has a set of
liquid crystal phases. In the opposite case one has a small
influence of the anisotropic part on the freezing transition
and, possibly, an orientational phase transition in solid state
which may affect or not the crystalline structure. All these
problems, including a rich phase diagram of liquid and molecular
crystal phases have been considered in the frame of DFT
approach.

Again in the case of hard-core molecules
(hard ellipsoids of
revolution (HER), hard spherocylindres (HSC) and hard
dumbells (HD)) one obtains very good agreement of DFT with
simulations. In this case the equivalent of HS potential is
(hard body potential HB) \begin{equation} u(r_{12}, {\bf
\Omega}_1, {\bf \Omega}_2) = \left\{ \begin{array}{rcl} \infty ,
r_{12}<D({\bf \hat r}_{12}, {\bf \Omega}_1, {\bf \Omega}_2)\\ 0,
\mbox {otherwise},\\ \end{array} 
\right.  \end{equation}
where $D(...)$ is the centre-
to-centre distance between the pair of molecules in contact for given
orientation of the pair.

3. In order to distinguish between purely geometrical
effects and bonding effects characteristic of real metals
Igloi et al. \cite{iglo87} have subtracted DCF of a HS fluid from the full
DCF of the metal which they determined from a perturbation theory. We may
note that for the HS system where the structure is determined solely by
the geometrical necessities of a closed packing, the structural
differences in the free energy (fcc and hcp) are extremely small. The
observed differences in metals are due to $\Delta c(k)$ which measures
the bonding effects.

4.
As an example of the application of quantun DFT theory of freezing
one can mention the paper by Rick et al. \cite{rimcha90}. They have
used this theory to calculate the phase diagram of ${}^{3}He$ and
${}^{4}He$ basing on second-order DFT and taking for the input the
correlation functions found from the recently developed quantum MC
technique.

All the investigations mentioned above bring to the conclusion that the
DFT of freezing is a good enough theory in the case of short-range
potentials or high pressures, when all potentials become more or less
hard body potentials. In the case of long-range potentials the
predictions of DFT theory differ essentially from real and computer
experiments. This can be demonstrated using as the example the
freezing of soft spheres, in which case
the fcc or the bcc phase was found to be a stable structure
independendently of $n$ but depending on the approximation used.

Recently in the frame of DFT a number of interesting results
were obtained for systems widely under investigation now both
experimentally and by use of computer simulations. Two such
systems are particularly "en vogue" now: colloidal suspensions
of polydisperse spheres and extr$\hat e$me potential systems
with isostructural transitions and the absence of liquid phase
as in $C_{60}$.

The geometrical nature of the theory defines its success in the
case of mixtures of hard core particles.  Mixtures exhibit phase
diagrams which are much richer than those of one component
systems including crystals substitutionally ordered or disordered.
The relative stability of these phases depends on the thermodynamic
conditions, concentration ratios and ratios of the sizes of atoms.
The freezing transition in polydisperse HS system is a very
interesting problem particularly because there exists now the
experimental realization of this simple theoretical model --
colloidal particles of different size (10-1000nm).

The DFT treatment of Barrat, Baus and Hansen
\cite{babaha8687} was the first to show that, starting
from monodisperse limit, the freezing transition of the fluid mixture into
a constitutionally disordered solid changes from a spindle type (at
s=0.94) to azeotropic type and (at s=0.92 /later 0.875) to
eutectic type  ($s=\sigma_1/\sigma_2$). Later the DFT approach
of \cite{babaha8687} was improved in \cite{zeox90} and
\cite{deas90}. The same results were obtained in simulations by
Kranendonk and Frenkel \cite{krfr91}.

The crystallization of the polydisperse hard sphere system with
continuous distribution of particle size (given by gamma-- or
Gauss distribution) was investigated by use of DFT approach
in a number of papers. As early as in 1986 in
\cite{poly86} (see also \cite{mrha88})
the critical value of polydispersity ($\approx 5-6\%$)
was obtained giving the limit to the possibility of
crystallization. Later the simulations  ~\cite{poly96}and the
real experiments confirm  the existence of such critical
dispersity (of a little greater value).

The other interesting experimental results ~\cite{coll}
on colloidal suspensions concern the phase diagram of
"big" spherical colloidal spheres in the liquid of very
small polymers. This systems was modelled through additional
attraction in the hard sphere potential. The range of the
attraction reflects the polymer size ~\cite{polha}.
If the polymer size decreases the temperature of the
liquid--gas transition decreases, too, and becomes
lower than the triple point temperature. Consequently,
the solid phase and only one fluid phase remain on the phase
diagram \cite{4,5,6,7}. If the range of the attraction
is decreasing more an isostructural polymorphic phase
transition appears in solid phase (in the simulations and in the
DFT approach ~\cite{ramena95,5,7,8,9,10}).
The main characteristics of phase transitions
and the phase diagrams from
the DFT calculations are in accordance with simulations
(~\cite{full93,fullek,full94,full96a,full96,mednew}).
Analogous results were obtained for hard spheres with the
Yukawa tail
or for adhesive and sticky spheres
\cite{laird90,pol,smha85,tb93,rice6}.

This problem -- the investigation of the attraction role
for the crystallization -- is a very interesting problem,
considered as early as by van der Waals.
The modern view on the problem can be found in
~\cite{week95,st96}.

To summarize the first part of the lecture, one can see that
the DFT of freezing is simple theory which works well
enough in the 3D case when one needs not to take into
account the fluctuations.

The properties of 2D crystals are quite different from the 3D
case. As we have mentioned above, in 30-th Peierls
has shown that 2D harmonical lattice cannot exist in the
thermodynamic limit:
the meansquare displacement in harmonic
crystals has logarithmic singularirity.
Later the absence of two--dimensional long--range translational
order was proved by Mermin ~\cite{merm68} with the use of well
known Bogoliubov inequalities ($1/q^2$ theorem) for the
correlators \cite{bogol1}.
In fact, in the paper by Mermin \cite{merm68}
mentioned above (where the absence of
crystallographic 2D order was proved), it was shown that the
true long range order -- the orientational order of bonds --
exists in 2D crystal. More precisely, it was shown that the
direction of the vector between any two neighbouring atoms at
finite temperature is the same as at $T = 0$.

The bond orientation correlation function
$<\vartheta({\bf r}_1)\vartheta({\bf r}_2)>$
remains finite at $r \rightarrow \infty$.
Here $\vartheta$ is the angle between local crystallographic
axis and some axis of ideal lattice. In continuum approximation
it has the form \begin{equation}
\vartheta(x,y)=\frac{1}{2}(\partial_x u_y-\partial_y u_x). \label{angle}
\end{equation}

Anisotropic liquids with bond--orientational order are
considered in  the phenomenological theories of melting,
developed by
Halperin and Nelson \cite{halpnel79} and Young \cite{young}
and based on the ideas of Kosterlitz and
Thouless \cite{kosthoul73}. This KTHNY theory predicts that the transition
may be fundamentally different from that observed in ordinary
three-dimensional systems. It was found that the transition between
two-dimensional solid and isotropic liquid can occur via two continuous
transitions corresponding, respectively, to dissociation of dislocations
and disclinations. The low-temperature solid phase is characterized by
algebraic decay of translational order
and true long--range bond-orienational order. Dislocations
unbind at a temperature $T_{\rm m}$ into a phase with shot-range
translational order, but with algebraic decay of
bond-orientational order. This intermediate phase is called the
hexatic phase. The properties of the hexatic phase are similar
to those of a nematic liquid crystal, except  that triangular
lattices melt into a phase with persistent sixfold, rather than twofold
order. Paired disclinations in the hexatic phase ultimately unbind
themselves, driving a second transition at a higher temperature $T_{\rm i}$
into an isotropic liquid.

The base of the KTHNY theory is the mechanism of breaking of
quasi--long--range order in 2D systems with continuous symmetry
developed by Beresinskii \cite{ber1,ber2} and by Kosterlitz and
Thouless \cite{kosthoul73,kost74}.
The wel known example of such system is the classical
$XY$ model with the Hamiltonian:
\begin{equation}
H=-\frac{J}{2}\sum_{<i\neq j>} {\bf S}_i{\bf S}_j \simeq \frac{J}{2}
\int\,d^2r\,(\nabla \varphi)^2.\label{xy}
\end{equation}
$\varphi$ is the angle between the vectors ${\bf S}_i$ and ${\bf
S}_j$ ($i$, $j$ - the nearest neighbours).

At low temperature the quasi--long--range order exists
characterized by an algebraic correlation decay
$<{\bf S}({\bf r}){\bf S}({\bf 0})> \propto r^{-\eta}$.
At higher temperature the correlations decay exponentially
$<~{\bf S}({\bf r}){\bf S}({\bf 0})> \propto \exp(-r/\xi)$.
The symmetry breaking occurs through the appearence of
free topological defects -- vortices:
$\oint\,(\nabla\varphi)\,dl =2\pi q$.

The transition temperature can be obtained simply from energetic
ballance using the fact that the vortex energy is (from
(\ref{xy}))
\[ E_{\nu}=\frac{J}{2}\int_0^L\,\frac{2\pi}{r}\,dr=J\pi\ln(L/a),\]
where $a$ is the lattice costant and $L$ - the system size.
The creation of a vortex changes the free energy:
$\Delta F=E_{\nu}-TS$, where the vortex entropy $S=2k_B
\ln(L/a)$. At the temperature $T\geq T_0=\pi J/k_B$
the value of
$\Delta F=(J\pi-2k_BT)\ln(L/a)$ becomes negative so that the
vortex creation becomes energetically profitable.
The Hamiltonian for vortex--vortex interaction is equivalent to
that of 2D Coulomb gas:
\begin{equation}
H_c=-\pi J\sum_{|{\bf r}-{\bf r}'|>a}q({\bf r})q({\bf r}')\ln\frac{|{\bf r}-
{\bf r}'|}{a}+E_c\sum_{\bf r}q^2({\bf r}). \label{hc}
\end{equation}

The physics of KT transition \cite{kosthoul73}
is the dissociation of vortex pairs in the presence of the
screening. The KT theory presents a renormgroup approach
to screening. The KT transition is a continuous transition
from low temperature quasi--long--range ordered phase to
high temperature disordered phase. The $T_{KT}$ is just that
obtained above from simple thermodynamical consideration.

Based on the KT theory the
KTHNY theory describes the 2D melting as two continuous KT
transitions. The first transition -- the dislocation pairs
unbinding, the second one -- the unbinding of disclination
pairs.

The description of the first transition in KTHNY theory is
based on the elastic Hamiltonian for 2D triangle lattice:
\begin{equation}
H_E=\frac{1}{2}\int\,d^2r\,\left[2\mu u_{ij}^2+\lambda u_{kk}^2\right],
\label{elast}
\end{equation}
where
\begin{equation}
u_{ij}=\frac{1}{2}\left[\frac{\partial u_i({\bf r})}{\partial r_j}+
\frac{\partial u_j({\bf r})}{\partial r_i}\right] \label{uij}
\end{equation}
and $\mu$ and $\lambda$ are the Lam\'e coefficients.

Free dislocations (analogs of the vortices in $XY$ model)
break the quasi-long-ranged periodic translational order and
cause the shear modulus $\mu$ to vanish.



The dislocation Hamiltonian has the form:
\begin{eqnarray}
H_{dis}&=&
-\frac{a_0^2K}{8\pi}\sum_{i\neq j}^M\left\{{\bf b}({\bf r}_i)
{\bf b}({\bf r}_j)\ln\frac{r_{ij}}{a}-\frac{({\bf b}({\bf r}_i){\bf r}_{ij})
({\bf b}({\bf r}_i){\bf r}_{ij})}{r_{ij}^2}\right\}+\nonumber\\
&+&E_d\sum_{i=1}^M\,{\bf b}^2({\bf r}_i),\label{hdis}
\end{eqnarray}
where $E_d$ is the core energy of dislocation.
\begin{equation}
K=\frac{4\mu(\mu+\lambda)}{2\mu+\lambda}.\label{K}
\end{equation}
The KT phase transition to the hexatic phase takes place at the
temperature
\begin{equation}
\frac{a_0^2K(T_m)}{k_BT_m} = 16\pi.\label{trpoint}
\end{equation}



Following Halperin and Nelson \cite{halpnel79},
the phenomenological orientational order parameter is
\begin{equation}
\psi({\bf r})=e^{6i\vartheta({\bf r})}, \label{pp}
\end{equation}
where $\vartheta({\bf r})$ is the bond orientation
(\ref{angle}).
The quasi-long-ranged orientational order at
$T>T_m$ is described in terms of algebraic correlation decay:
\begin{equation}
<\psi^*({\bf r})\psi(0)>\propto r^{-\eta_6(T)}. \label{hhcorr}
\end{equation}
The Hamiltonian for the hexatic phase has the form
\cite{halpnel79}:
\begin{equation}
H_A=\frac{1}{2}K_A(T)\int\,d^2r\,(\nabla \vartheta({\bf r}))^2, \label{hhex}
\end{equation}
The Frank constant $K_A(T)$ is connected to $\eta_6$ through
the equation:
\begin{equation} \eta_6(T)=\frac{18 k_BT}{\pi
K_A(T)}. \label{eta6} \end{equation}

In the solid phase the disclinations are tightly bound into
pairs (the dislocations), however in the hexatic phase
the interaction is screened by dislocations and so
occurs to be logarithmic
\cite{halpnel79,nem1}, and so for the disclination
Hamiltonian in hexatic phase we obtain the 2D Coulomb
gas like form again \cite{halpnel79,nelson82}:
\begin{equation}
H_{disc}=-\frac{\pi K_A(T)}{36}\sum_{{\bf r} \neq {\bf r}'}s({\bf r})s({\bf r}')
\ln\frac{|{\bf r}-{\bf r}'|}{a}+E_{cd}\sum_{\bf r}s^2({\bf r}),
\label{hdisc}
\end{equation}
where $E_{cd}$ is the core energy of disclination, $s({\bf r})=1$
for the atom with 7 nearest neighbours,
$s({\bf r})=-1$ for 5 nearest neighbours.

The second KT transition -- the disclination pairs unbinding --
takes place at the temperature \cite{halpnel79} $T_i>T_m$:
\begin{equation}
T_i=\frac{\pi K_A(T_i)}{72 k_BT_i}.
\label{Ti}
\end{equation}

The KTHNY picture is the following.
At $T=0$ there exist long-range translational and orientational orders,
however, at $T>0$ the long-range translational order is transformed into
the quasi-long-range one due to smooth phase fluctuations, the
orientational order remaining unchanged. At the melting temperature
$T_{\rm m}$, the quasi-long-range order is destroyed by the singular phase
fluctuations (dislocations).
The appearance of free dislocations means that the system ceases
to offer any resistance to shear ($\mu = 0$),
i.e., it becomes a liquid.
It should be emphasized that the amplitude of
the order parameter $\rho_{\bf G}$ does not become equal to zero
at $T_{\rm m}$, but it does at the some mean-field temperature $T_{\rm MF}$,
which can be determined by equating the free energies of the solid and
liquid phases as the functionals of the local density.
(It is important to note that it is
this point $T_{\rm MF}$ that has been considered by Ramakrishnan
\cite{rama82} as the melting point.)
The second
transition into an isotropic liquid takes place at a higher
temperature $T_{\rm i}$ .

This theory has strong support from experiments with electrons on helium
\cite{gram,electrons}, computer simulations of the 2D electron systems
\cite{morf79,lozov85},elasticity simulations \cite{saito}, experiments with
polystyrene spheres \cite{mu} (however, the topological defects in this
case are complex and are not consistent with a simple KTHNY picture of
melting).

A conventional first-order transition between the two-dimensional solid
and isotropic liquid is also a possibility. Several theories predict
a single first-order melting transition in 2D
\cite{rama82,chui83,klei83,ryz91,ryzjetp91,lozfar}.
The sum of the experimental evidence on adsorbed atoms is weighted against
an interpretation in terms of KTHNY melting and seems to show a weak
first-order transition \cite{ads}. Most simulations studies of 2D melting
indicate that the strictly two-dimensional hard-core potential systems
melt via first-order transition (see, for example,
\cite{strandburg88,strandburg92} and references therein), as do studies
of systems interacting with intermediate strength potentials
\cite{lozov85,interm}.

Therefore, we expect that the melting behavior may depend crucially on the
interaction potential, and the first-order character of the transition is
weakened as the potential is softened.

The simple physical picture may be painted in order to illustrate this
possibility \cite{ryta95,ryta95a}.
As mentioned above, there is an analogy between the hexatic phase
and a nematic liquid crystal, but the role of the rodlike molecules is
played by the hexagon clusters consisting of an atom with its nearest
neighbors. Ordering of these hexagons is possible only if the range
of the interparticle interaction is large enough to provide the interaction
between clusters. Therefore, the interparticle interaction should extend
at least over several interparticle distances, and hexatic phase could
exist for long-range potentials, but not for shot-range ones. However,
only first-principles study could put these heuristic reasonings on firmer
footing.

The microscopic theory of 2D melting can be obtained as a
generalization of the DFT theory of freezing described above.
We developed an approach appropriable to 2D melting
\cite{ryta87,ryta88,ryz89,ryz90,ryta91,ryta92,ryta93,%
ryta93a,ryta95,ryta95a}.  Our approach can predict, basing on
the knowledge of interparticle potential, which scenario is to
be realized: 1) $T_{\rm m}< T_{\rm MF}$, the system melts by
means of two continuous transitions of KT type; 2) $T_{\rm MF}<
T_{\rm m}$, the system melts by means of a first-order
transition.

Our approach differs from the standard DFT theory of freezing in two
main points:

First, we permit
the Fourier coefficients $\rho_{\bf G}({\bf r})$ of the
one-particle distribution function expanded in a Fourier series
in reciprocal-lattice vectors $\{ {\bf G} \}$:
$$ \rho({\bf r}) = \sum_{\bf G} \rho_{\bf G}({\bf r}) e^{i{\bf G r}} $$
to fluctuate - to vary slowly over
distances of order $G^{-1}$ and to have the amplitude and phase:
$$\rho_{\bf G}({\bf r}) = \rho_{\bf G} e^{i{\bf G u}({\bf r})}.$$
Here ${\bf u}({\bf r})$ is the displacement field.

Second, we permit the liquid to be anisotropic: we consider as possible
the existence of a phase with constant density but angular dependent
two-particle distribution function
$F_2({\bf r}_1 -{\bf r}_0) \neq g(r_{10})$.

These two points of generalization define the two new order parameters:
the fluctuating $\rho_{\bf G}({\bf r})$ and the Fourier coefficients
characteristic for the broken symmetry of the function
$F_2({\bf r}_1 -{\bf r}_0)$.
Our approach again is based on the Eq.(\ref{main}) of previous section.

Let us describe in details the
microscopic characteristics of a hexatic phase.

The relative spatial distribution of pairs of particles is characterized
by the function
$F_2({\bf r}_1|{\bf r}_0) = F_2({\bf r}_1 -{\bf r}_0)$.
The vector
${\bf r}_1-{\bf r}_0$
defines the direction of the bond between the molecules at the points
${\bf r}_1$ and ${\bf r}_0$. In the ordinary isotropic liquid the
nearest neighbouring of a given molecule (the first coordination
sphere) has a definite local symmetry, which can be characterized by the
set of bond directions. The local structure of the liquid in the
neighbourhood of a molecule at the point ${\bf r}_0'$ is characterized
by the bond directions ${\bf r}'={\bf r}_2 - {\bf r}_0'$. It occurs
that if the point ${\bf r}_0'$ is at sufficiently large distance from
${\bf r}_0$ then there is no correlation between the directions
${\bf r}={\bf r}_1 - {\bf r}_0$ and
${\bf r}'={\bf r}_2 - {\bf r}_0'$.  In this case after the averaging over
the system as a whole the pair distribution function transforms into
the RDF and the equation (\ref{main}) for $s=1$ has the
solution
$F_2({\bf r}_1-{\bf r}_0) = g(|{\bf r}_1-{\bf r}_0|)$,
which corresponds to ordinary isotropic liquid.

When we approach the anisotropic liquid phase the long--ranged
correlations between the bond directions
${\bf r}$ and ${\bf r}'$ do appear and the averaged two--particle
distribution function depends on the bond direction now.

In the vicinity of the transition one can write
\begin{equation}
F_2({\bf r}_1, {\bf r}_0) = g(|{\bf r}_1-{\bf r}_0|) (1+f({\bf r}_1-
{\bf r}_0)) \label{hex}
\end{equation}
where $f({\bf r}_1|{\bf r}_0)$  has the symmetry of the local
neighbourhood of the particle at ${\bf r}_0$.
In the two--dimensional case
\begin{equation}
f({\bf r}_1|{\bf r}_0)=f(a_0, {\bf r}_0, \varphi). \nonumber
\end{equation}
Here $а_0=|{\bf r}_1-{\bf r}_0|$, $a_0$ is the radius of the first
coordination sphere, and
$\varphi$  is the angle of the vector ${\bf a}_0$. The function
$f$ may be expanded in a Fourier series
\begin{equation}
f(a_{0},{\bf r}_{0},\varphi)=\sum_{m=-\infty}^{\infty}
f_{m}(a_{0},{\bf r}_{0})
e^{im\varphi}. \label{furier}
\end{equation}

The Fourier coefficients define the order parameters. These parameters
become nonzero at the  temperature $Т_{MF} \leq T_h$
defined by the bifurcation condition, that is by the eigenvalue of
the linearized (relative to $f$) eq. (\ref{main}) with $s=1$.

At the same time, when one approaches the line defined by the
bifurcation condition, the correlation radius for the orientation
fluctuations of the pair distribution function diverges. This fact
can be shown with the use of the gradient expansion technique in
the case of the equation  (\ref{main}) for $s=3$, if we write the long
range part of the correlator using the principle of a weakening of
correlations (\cite{NNB2}) as:
\begin{equation}
F_4({\bf r}_1, ..., {\bf r}_4)=g(|{\bf r}_1 - {\bf r}_2|)
g(|{\bf r}_3 - {\bf r}_4|) (1+  f_4({\bf r}_1, ..., {\bf r}_4))
\label{f4}
\end{equation}
\[ f_4({\bf r}_1, ..., {\bf r}_4)= f_4(r, R, \rho, \varphi_1, \varphi_2).
\] Here $\varphi_1$ is the angle between the vector ${\bf r}={\bf
r}_1-{\bf r}_2$ and the axis ${\bf R}={\bf r}_2-{\bf r}_3$, $\varphi_2$ is
the angle between the vector ${\bf \rho}={\bf r}_3-{\bf r}_4$  and the
same axis.
We have
$f_4(r, R, \rho, \varphi_1, \varphi_2) \rightarrow 0$ when
$R \rightarrow \infty$ .

The function $f_4$ can be presented as a Fourier series
\begin{equation}
f_4(r, R, \rho, \varphi_1, \varphi_2)=\sum_{p,q} f_{pq}(r, R, \rho)
\exp(ip\varphi_1+iq\varphi_2). \nonumber
\end{equation}

The asymptotics of the solution is of the form
\[ f_{pq} \propto e^{-R/\xi_{pq}}. \] Here the correlation length
$\xi_{pq}$ is characterized by the properties of the isotropic liquid and
diverges in the vicinity of the bifurcation line. The asymptotic behaviour
of the correlation function $f_{6,6}$ , derived in our approach,
(far away from the transition) can be
compared with that of the phenomenological KT theory, thus giving the
microscopic expression for Frank constant \cite{ryta92}:
\begin{eqnarray}
K_{ A}(T)&=&648 k_{ B} T a_0^2 |f_6|^2  [\Gamma_6(a_0,a_0)
-1/2(\Gamma_5(a_0,a_0)+ \Gamma_7(a_0,a_0))], \label{KA0}\\
\Gamma_m(r_1,r_2)&=&\frac{1}{2\pi}\int_0^{2\pi} d\varphi\,
\Gamma(r_1,r_2,\varphi) e^{-im\varphi} \nonumber
\end{eqnarray}
where $a_0^2=2/(\sqrt{3} \rho)$.
The function $\Gamma(r_1,r_2,\varphi)$ was introduced in our
papers \cite{ryta87,ryta88,ryz89,ryz90} on bond orientational
order and has the form
\begin{eqnarray}
\Gamma(r_1,r_2,\varphi)&=&\sum_{k \geq 1}\frac{\rho^k}{(k-1)!}
\int\, S_{k+1}({\bf r}_1,...,{\bf r}_{k+1}) \nonumber\\ &\times
& g(|{\bf r}_3-{\bf r}_0|)\cdots g(|{\bf r}_{k+1}-{\bf r}_0|)
d^2 r_3 \cdots d^2 r_{k+1}, \label{Gamma} \end{eqnarray} where,
as earlier, $S_{k+1}({\bf r}_1,...,{\bf r}_{k+1})$ is the
irreducible cluster sum of Mayer functions connecting (at least
doubly) $k+1$ particles, $r_1=|{\bf r}_1-{\bf r}_0|, r_2=|{\bf
r}_2-{\bf r}_0|$, and $\varphi$ is the angle between the vectors
${\bf r}_1$ and ${\bf r}_2$.

In the spirit of DFT we expand the function
(\ref{Gamma}) in a functional Taylor series in powers of
$h(r)=g(r)-1$:
\begin{eqnarray}
\Gamma(r_1,r_2,\varphi)&=&\rho\left( c^{(2)}(|{\bf r}_1-{\bf r}_2|)+
\sum_{n=1}^{\infty}\frac{\rho^n}{n!} \int\,
c^{(n+2)}({\bf r}_1,...,{\bf r}_{n+2}) \right.\nonumber \\
& \times&\left. h(|{\bf r}_3-{\bf r}_0|)\cdots h(|{\bf r}_{n+2}-{\bf r}_0|)
d^2 r_3 \cdots d^2 r_{n+2}\right). \label{Gam_e}
\end{eqnarray}

In the hypernetted chain approximation we can drop
in the expansion (\ref{Gam_e}) all terms with $n\geq3$
and obtain
\begin{equation}
\Gamma(r_1,r_2,\varphi)=\rho c^{(2)}(|{\bf r}_1-{\bf r}_2|). \label{app}
\end{equation}

In this approximation, Eq. (\ref{KA0}) has the form
\begin{equation}
K_{ A}(T)=648 k_{ B} T a_0^2 |f_6|^2  [c_6(a_0,a_0)
-1/2(c_5(a_0,a_0)+
c_7(a_0,a_0))], \label{KA}
\end{equation}
where
\[c_{ m}(a_0,a_0)=\frac{1}{2\pi}\int_0^{2\pi}d\varphi c^{(2)}(\sqrt{2}a_0
(1-\cos\varphi)^{1/2};\hat{\rho}) e^{-im\varphi} .\]

Now the equation for the absolute value
of the order parameter is \cite{ryta87,ryta88,ryz89,ryz90}:
\begin{equation}
f_6=\frac{\int_0^{2\pi}\,d\varphi\,\cos 6\varphi\,
\exp[12 c_6(a_0,a_0)\,\cos(6\varphi) f_6]}
{\int_0^{2\pi}\,d\varphi\,\exp[12 c_6(a_0,a_0)\,\cos(6\varphi) f_6]}.
\label{f6}
\end{equation}
(We have made use of the Eq.(\ref{main}) for the function
 $ F_2({\bf r}_1|{\bf r}_0)$).
As in the case of standard DFT this equation corresponds to the
minimum of the free energy functional.

The phase transition from the hexatic phase to the 2D solid phase
can be analyzed in a similar way. In this case the correlator
which diverges is the density--density correlation function, the
density Fourier components being the order parameters $\langle
\rho_{\bf G}({\bf r}) \rho_{\bf G}({\bf 0})\rangle \propto
h_{\bf G}({\bf r}); \,\, h_2({\bf r})= \sum_{\bf G} h_{\bf
G}({\bf r}) e^{i{\bf G r}}$, where ${\bf G}$ are the reciprocal
lattice vectors.  \begin{equation}
\rho({\bf r})= \rho F_1({\bf r}) =\rho_0
+ \Delta \rho({\bf r})= \rho_0 + \sum_{\bf G} \, \rho_{\bf G}({\bf r}) \,
e^{i{\bf G r}}
\label{cryst}
\end{equation}
\[ \rho_{\bf G}({\bf r}) =
|\rho_{\bf G}| e^{i{\bf u(r)}} \]

In both cases the solution with broken symmetry appears at the point
where the correlation decay changes its character.

The microscopic expressions for the nonrenormalized Lam\'e
coefficients were obtained in
\cite{ryta92} by comparing the long-range behaviour of our
correlator with the asymptotic behavior of the order parameter
correlator from the phenomenological elastic energy
\cite{halpnel79}. For short-range potentials, expressions were
obtained in \cite{ryta92} for the elastic moduli which
correspond to fluctuations with wave vector equal to the
smallest reciprocal lattice vector. Generalization to the case
of an arbitrary number of inverse lattice vectors leads to the
following expression for the Lam\'e coefficients:
\begin{eqnarray}
\mu= && \frac{k_{ B} T}{16 \rho} \sum_G \, \rho_G^2 m_G G^2 (\gamma_G+
2\delta_G)
\label{mu},\\
\lambda= &&  \frac{k_{ B} T}{16 \rho} \sum_G \,
\rho_G^2 m_G G^2 (\gamma_G- 6 \delta_G)
\label{lambda},
\end{eqnarray}
where
\begin{eqnarray}
\gamma_G =&& 2 \pi \rho \int \,r^3d r\,c^{(2)}(r;\hat{\rho}) J_0(Gr)
\label{gamma},\\
\delta_G =&& 2 \pi \rho \int \,r^3d r\,c^{(2)}(r;\hat{\rho}) J_1(Gr)/(Gr),
\label{delta}
\end{eqnarray}
$J_0(x)$ and $J_1(x)$ are the Bessel functions, and $m_G$ is the
number of reciprocal lattice vectors with the same length.

In the case of the long-range Coulomb interaction, an additional
term arises in the elastic Hamiltonian \cite{lozov91},
which makes the effective modulus $\lambda$ diverge,
$\lambda = \infty$, but the expression (\ref{mu})for
$\mu$ remains to be valid. The modulus
$K$ takes the form:
\begin{equation} K=4
\mu a_0^2/k_{ B} T.  \label{kel} \end{equation}

In \cite{ryta95} the authors present
the first-principles estimates for the stability
limits of the solid and hexatic phases  for the 2D electron system and
the system of hard disks
which are two opposite cases of the $1/r^n$ potential. In \cite{ryta95a}
2D system of vortices in a superconducting film interacting via a
potential which is even softer than in 2D electron system was
considered. The transition temperatures $T_{\rm m}$ and $T_{\rm
i}$ were obtained from the KT theory using microscopic
expressions for elastic and Frank moduli and were compared with
the value of $T_{\rm MF}$ from the standard DFT theory.

1. Beginning with the classic paper by Alder and
Wainwright \cite{alder} it has been assumed that a system of
hard disks melts by means of the first order transition.
However, only recently have convincing proofs of this fact been
obtained \cite{strandburg92}. The standard DFT approach
gives good qualitative and quantitative agreement with
the results of computer simulations. In the simpliest version
we obtain $\rho_{ S}=0.933$ (to be compared with 0.921 from
simulations \cite{barker}). At the transition point we have
$K/16\pi=6.29$ which is much larger than the value $K/16\pi=1$,
at which the dissociation of dislocation pairs takes place.

However, an even more convincing argument in favor of the
first order phase transition is provided by the analysis
of the possible existence of hexatic phase. Using the equation
(\ref{KA}) we obtain for the Frank constant for the system of
hard disks $K_{A}(T) < 0$ at all density, so that the hexatic
phase cannot exist.

2. In the case of 2D classical Wigner crystal we have obtained
the two-stage scenario of melting. In terms of the dimensionless
parameter
$\Gamma = (\pi \rho)^{1/2} e^2/(k_{ B}T)$ our results are:
$\Gamma_{ MF}=21.58$, $\Gamma_{ m}^{ KT}=80.15$
and $\Gamma_{ i}^{ KT}=24.5$. However,
the renormalization of parameters depends strongly on the
unknown values of the core energies of disclinations and
dislocations and this can change the result.

3. In thin superconducting film the interaction energy
of two vortices located at the points
${\bf r}_i$ and ${\bf r}_j$
($r_{ij}= |{\bf r}_{i}-{\bf r}_{j}| \gg \xi$)
has the form \cite{pearl}
\begin{equation}
\begin{array}{lll}
\rule[-10pt]{0pt}{20pt}%
\Phi(r_{ij})=&\frac{\varphi_{0}^{2}}{8\pi \Lambda} \left[ H_{0}
\left(\frac{r_{ij}}{\Lambda} \right) - Y_{0}\left( \frac{r_{ij}}{\Lambda}
\right) \right] \\

\rule[-10pt]{0pt}{20pt}%
\Phi(r_{ij}) \approx &- \frac{\varphi_{0}^{2}}{4\pi^{2}\Lambda} \ln
\left( \frac{r_{ij}}{\Lambda} \right) & r_{ij} \ll \Lambda \\

\Phi(r_{ij}) \approx &\frac{\varphi_{0}^{2}}{4\pi^{2}r_{ij}} &
r_{ij} \gg \Lambda \end{array}
\label{pot}
\end{equation}
where $\Lambda(T)=2 \lambda_B^2(T)/d$
is the effective penetration depth,
$d$ is the film thickness, $\lambda_B$ is the bulk penetration
depth, $\varphi_0=hc/2e$ is the flux quantum,
$H_0(x)$ - the Struve function and $Y_0(x)$ - the Neumann
function. The potential (\ref{pot}) is long-ranged (it is
even "softer" than the Coulomb potential
$1/r$) and one might therefore expect that vortex lattice will
melt through two continuous transitions.
As to real and computer experiments, not only the type of the
transition, but even the very existence of vortex lattice
melting, remains an open question. This system was investigated
in details by the authors in
\cite{ryta93b,ryta94,irt1,irt2,irt3}.

On the basis of our approach we have made the calculations
for the specific case of a niobium film of thickness
$20 \AA$, investigated experimentally in Ref.
\cite{hsukapit92}.
The region of the
hexatic phase is very well pronounced on the H-T phase diagram.
If the film thickness increases this region narrows (see figures in
\cite{ryta95a}).

\smallskip

The authors thank D.Yu.Irz, Yu.L.Klimontovich, N.M.Plakida and
S.M.Stishov for helpfull discussions and valuable comments and
RFBR for financial support (grant N 96-02-16211).


\end{document}